\documentclass{cmspaper}
\usepackage{graphicx,epsfig,rotating,multirow,amssymb,url}
\begin{document}
\newcommand{\nc}{\newcommand}
\nc{\lsim}{\mbox{\raisebox{-.6ex}{~$\stackrel{<}{\sim}$~}}}
\nc{\gsim}{\mbox{\raisebox{-.6ex}{~$\stackrel{>}{\sim}$~}}}
\nc{\esim}{\mbox{\raisebox{-.6ex}{~$\stackrel{-}{\sim}$~}}}
\nc{\beq}{\begin{equation}}
\nc{\eeq}{\end{equation}}

\catcode`@=11
\def\citer{\@ifnextchar
[{\@tempswatrue\@citexr}{\@tempswafalse\@citexr[]}}

% \citer as abbreviation for 'citerange' replaces the ',' by a '--'
%

\def\@citexr[#1]#2{\if@filesw\immediate\write\@auxout{\string\citation{#2}}\fi
  \def\@citea{}\@cite{\@for\@citeb:=#2\do
    {\@citea\def\@citea{--\penalty\@m}\@ifundefined
       {b@\@citeb}{{\bf ?}\@warning
       {Citation `\@citeb' on page \thepage \space undefined}}%
\hbox{\csname b@\@citeb\endcsname}}}{#1}}
\catcode`@=12
 
%%%%%%%%%%%%%%%%%%%%%%%% DEFINITIONS
\newcommand{\ra}{\rightarrow}
%%%%%%%%%%%%%%%%%%%%%%%% TITLE PAGE

%==============================================================================
% title page for few authors

\begin{titlepage}

% select one of the following and type in the proper number:
%\cmsnote{2004/027}
%  \internalnote{2003/014}
%  \conferencereport{2005/000}

\date{12 November 2004}

\begin{flushright}
CMS NOTE 2004/027
\end{flushright}

\vspace*{2.0cm}

%\title{Estimate of tan$\beta$ measurement precision in H/A~$\ra\tau\tau$ and  
%  H$^{\pm} \ra \tau\nu$ in CMS}
  \title{Measurement of the H/A~$\ra\tau\tau$ cross section and possible 
         constraints on tan$\beta$} 

  \begin{Authlist}
     ~R.~Kinnunen\Iref{hip},
     ~S.~Lehti\Iref{hip},
     ~F.~Moortgat\Iref{belg},
     ~A. ~Nikitenko\IAref{impe}{a},
     ~M. ~Spira\Iref{PSI}
  \end{Authlist}
  \Instfoot{hip}{Helsinki Institute of Physics, Helsinki, Finland}
  \Instfoot{belg}{CERN, PH Department, Geneva, Switzerland} 
  \Instfoot{impe}{Imperial College, University of London, London, UK}
  \Instfoot{PSI}{Paul Scherrer Institut, CH-5232 Villigen PSI, Switzerland}
  \Anotfoot{a}{On leave from ITEP, Moscow, Russia}

  \begin{abstract}

The achievable precision of the cross section times branching ratio
measurement from the event rates is
estimated for the MSSM H/A~$\ra\tau\tau$ decay
in the associated production process gg~$\ra \rm b \bar{\rm b}\rm H/A$ 
at large tan$\beta$ in CMS. This work demonstrates that the above production
and decay process exhibit a large sensitivity to tan$\beta$ and thus add
as a significant observable to a global fit of the SUSY parameters. To
illustrate this potential an example is given concerning the achievable
tan$\beta$ determination accuracy that could be reached from the event rates
and for a given set of SUSY parameters and uncertainties.
 
  \end{abstract} 

% if needed, use the following:
%\conference{Presented at {\it Physics Rumours}, Coconut Island, April 1, 2005}
%\submitted{Submitted to {\it Physics Rumours}}
%\note{Preliminary version}
  
\end{titlepage}

\setcounter{page}{2}%
%==============================================================================

%\input{tanbeta_text_v23cmsnote}
\section{Introduction}

The Higgs mechanism is a cornerstone of the Standard Model (SM) and its
supersymmetric extensions. Therefore, the search for Higgs bosons is one of
the top priorities at future high-energy experiments.  Since
the Minimal Supersymmetric extension of the Standard Model (MSSM)
requires the introduction of two Higgs doublets in order to preserve
supersymmetry, there are five elementary Higgs particles, two CP-even
(h,H), one CP-odd (A) and two charged ones (H$^\pm$). At lowest
order all couplings and masses of the MSSM Higgs sector are determined by two
independent input parameters, which are generally chosen as
tan$\beta=v_2/v_1$, the ratio of the vacuum expectation values of the two
Higgs doublets, 
and the pseudoscalar Higgs boson mass m$_{\rm A}$. At leading order 
(LO) the light
scalar Higgs boson mass m$_{\rm h}$ has to be smaller than the Z boson mass m$_{\rm Z}$.
Nevertheless the upper bound is significantly enhanced by radiative
corrections, the leading part of which grows with the fourth power of
the top mass $\rm m _{\rm t}$ and logarithmically with the stop masses. 
Therefore, the upper bound on $\rm m_{\rm h}$ is increased to 
m$_{\rm h}\lsim 135$ GeV$/c^2$ when one-loop and dominant two-loop corrections 
are included \cite{mssmrad}. 
The negative direct searches
for the Higgsstrahlung processes $e^+e^-\to\rm Zh/ZH$ and the associated
production $e^+e^-\to\rm Ah/AH$ yield lower bounds of m$_{\rm h,H} > 91.0$
GeV$/c^2$ and m$_{\rm A} > 91.9$ GeV$/c^2$. The range $0.5 < $tan$\beta < 2.4$
in the MSSM is excluded for m$_{\rm t}=174.3$ GeV$/c^2$ by the Higgs searches at
the LEP2 experiments \cite{lep2}.

Thus, one of the most important parameters to be determined in the
Minimal Supersymmetric Standard Model as well as in a general
type-II Two-Higgs Doublet Model (2HDM) is tan$\beta$.  In the MSSM
tan$\beta$ plays a crucial role, since it characterizes the relative
fraction of the two Higgs boson doublets contributing to the electroweak
symmetry breaking.  Consequently, it enters in all sectors of the
theory.  For small tan$\beta$, it may be possible to determine the value
of tan$\beta$ within the sfermion or neutralino sector \cite{baer,
lali}. For large tan$\beta$ this method has not been found to be
effective. In this regime, however, there are good prospects to measure the
value of tan$\beta$ by exploiting the Higgs sector \cite{gunion}.

At large tan$\beta$ neutral and charged Higgs boson production is
dominated by the bremsstrahlung processes $\rm gg\rightarrow b\bar{\rm
b}H/A$ and gb~$\ra$ tH$^{\pm}$. The dominant parts
of the production cross sections are proportional to
tan$^2\beta$. For the heavy scalar MSSM Higgs boson H this
behavior is valid within 1\% for tan$\beta\gsim 10$, if the
pseudoscalar mass m$_{\rm A}$ is larger than about 200 GeV$/c^2$, while for
m$_{\rm A} > 300$ GeV$/c^2$ it is already satisfied for tan$\beta\gsim 5$. 
Due to this feature the uncertainty on the tan$\beta$ measurement is only
half of the uncertainty on the rate measurement. In the MSSM the
supersymmetric loop corrections introduce an additional tan$\beta$
dependence to the cross section \cite{deltamb}, but they can be absorbed
in an effective parameter tan$\beta_{\rm eff}$, since the dominant terms
which are enhanced by tan$\beta$ correspond to emission and re-absorption
of virtual heavy supersymmetric particles at the bottom Yukawa vertex,
which are confined to small space-time regions compared to QCD
subprocesses involving massless gluons. The sub-leading terms are small.
The dominant terms are universal contributions to the bottom Yukawa
coupling \cite{deltamb}.  This implies that the method described below
determines this effective parameter tan$\beta_{\rm eff}$ in the MSSM
(for simplicity, it is denoted everywhere in the rest of the text 
as tan$\beta$). 
The extraction of the fundamental tan$\beta$ parameter requires
additional knowledge of the sbottom and gluino masses as well as the
Higgs boson mass parameter $\mu$. These corrections are in general absent 
in a 2HDM so that in these models the extracted value belongs to the 
fundamental tan$\beta$ parameter.
%The following analysis is valid for the 2HDM and the general MSSM, if
%Higgs boson decays into supersymmetric and other non-standard particles
%are kinematically forbidden or suppressed.
The H/A~$\ra\mu\mu$ \cite{bellucci}
and H/A~$\ra\tau\tau$ decay channels have been identified as the most
promising ones for the search of the heavy neutral MSSM Higgs boson H and A
at large tan$\beta$. The final states $\rm e\mu$, $\ell\ell$ ($\rm
\ell\ell=e\mu, ee,\mu\mu$) \cite{2lepton}, lepton+jet \cite{hljet} and
two-jets \cite{2jets} have been investigated for the H/A~$\ra\tau\tau$
decay mode. In this study the $\rm m _{\rm h} ^{\rm max}$
scenario \cite{carena} have been chosen with the following MSSM 
parameters: SU(2) gaugino mass M$_2$ = 200~GeV/$c^2$, $\mu$ =
300~GeV/$c^2$, gluino mass M$_{\tilde{\rm g}}$ = 800~GeV/$c^2$, 
SUSY breaking mass parameter M$_{\rm SUSY}$ = 1 TeV/$c^2$ and stop mixing
parameter X$_{\rm t}$=$\sqrt{6}$ (X$_{\rm t}$=A$_{\rm t}$-$\mu$cot$\beta$).
The top mass is set to 175~GeV/$c^2$. The Higgs boson decays to SUSY 
particles are allowed.

In this work the Next-to-Leading Order (NLO)
cross section and the branching ratio theoretical uncertainty; and the 
luminosity and statistical uncertainties are taken into account. 
The precision on the mass 
measurement in the H/A$\rightarrow\tau\tau$ channel is estimated and  
included when trying to determine the precision on the tan$\beta$ 
measurement. The uncertainty on the background as well as the uncertainty on 
the signal selection efficiency are included as described in Section 4.

The precision that can be achieved in the measurement of the cross section
times branching ratio from the event rate within the MSSM for the 
H/A~$\ra\tau\tau$ decay channel in the associated production process 
$\rm g \rm g \rightarrow \rm b \bar{\rm b}H/A$ at large tan$\beta$ is 
estimated in this study. The work also demonstrates that the above production
and decay process exhibits a large sensitivity to tan$\beta$ becoming a 
significant observable to be included in a global fit of the SUSY parameters.
To illustrate this potential an example is given concerning the achievable 
tan$\beta$ measurement accuracy that could be reached for the above set of
SUSY parameters.

The note is organized as follows: Section 2 describes the event selection
and the expected discovery reach. The measurement uncertainty on the 
cross section is discussed in Section 3. Section 4 presents the evaluation 
of the tan$\beta$ measurement uncertainty; and in Section 5 the 
total uncertainty on the tan$\beta$ measurement is summarized. Finally,
Section 6 is devoted to the conclusions.

\section{Event selection and expected discovery reach}

This section starts with the comparison between the theoretical NLO cross
section and the PYTHIA cross section since PYTHIA has been used for the 
generation of the hard processes. Then the event selection is
discussed, and finally, the expected discovery reach is presented.

\subsection{NLO cross section and PYTHIA cross section comparison}

Higgs boson production in the $\rm gg \rightarrow \rm b \bar{\rm b}H/A$
process was obtained with the PYTHIA \cite{pythia} two-to-three
processes 181 and 186 and with the PYTHIA6.158 default values for the
parton distribution functions and the renormalization and factorization
scales. No cut on the transverse momentum of the $\rm b$ quarks has been
applied at the generation level but $\rm E_{\rm T}^{\rm jet}>$~20 GeV is
used for the b-jet identification in the event analysis.  Therefore it
is important to know how well PYTHIA describes the $\rm p_{\rm T}$
spectrum of the $\rm b$ quarks compared to the NLO calculations 
\cite{hep-ph/0309204}
in order to estimate how well the efficiency of the event selection can
be determined.
The comparison was made for the SM process $\rm gg\ra b\bar{b}h$ 
(PYTHIA process 121)
with a Higgs boson mass of 120 GeV/$c^2$.
The PYTHIA and the NLO cross sections are compared in
Table~\ref{table:cross_section} as a function of a cut on the transverse
momentum of the $\rm b$ quark with highest $\rm p_{\rm T}$. In PYTHIA as well
as in the NLO calculations the $\rm b$ quark momentum was taken after gluon
radiation. The total PYTHIA cross sections (p$_{\rm T}>0$) were
normalized to the total NLO cross sections. The agreement between the
PYTHIA and the NLO values turns out to be at the level of 5--10\%. The
statistical uncertainties of the PYTHIA cross sections are shown, too.
For completeness the PYTHIA LO cross sections are also compared to the
corresponding theoretical LO calculation (the lower two rows in
Table~\ref{table:cross_section}). In this case the PYTHIA $\rm b$ quark was
taken before gluon radiation. Good agreement within 1--2 \% has been
obtained.

\begin{table}[h]
%  \vskip 0.1 in
  \caption{Comparison of the NLO and LO cross sections to the PYTHIA
  cross sections as a function of the cut on the transverse momentum of
  the b quark with highest $\rm p_{\rm T}$. The total PYTHIA cross
  sections ($\rm p_{\rm T}>$ 0) are normalized to the corresponding
  NLO(LO) cross sections.}
  \label{table:cross_section}
  \centering
  \vskip 0.2 in
  \begin{tabular}{|l|c|c|c|c|c|c|}
    \hline
     p$_{\rm T}$ cut & 0 GeV/$c$ & 10 GeV/$c$ & 20 GeV/$c$ & 30 GeV/$c$ & 40 GeV/$c$ & 50 GeV/$c$ \\
    \hline
    $\sigma_{\rm NLO}$ (pb)    & 734 & 507 &  294 & 173 & 106 &  68 \\
    $\sigma_{\rm PYTHIA}$ (pb) & 734 %$\pm$ 0 
                               & 523 $\pm$ 3
                               & 275 $\pm$ 3
                               & 156 $\pm$ 3
                               &  92 $\pm$ 2
                               &  60 $\pm$ 2 \\
    \hline
    $\sigma_{\rm LO}$ (pb)     & 528 & 393 &  241 & 152 & 102 &  71 \\
    $\sigma_{\rm PYTHIA}$ (pb) & 528 %$\pm$ 0
                               & 407 $\pm$ 2
                               & 245 $\pm$ 3
                               & 154 $\pm$ 2
                               & 101 $\pm$ 2
                               &  70 $\pm$ 2 \\
    \hline
  \end{tabular}
 \vskip 0.1 in
\end{table}

\subsection{Event selection}

The event selection for the two-lepton (e$\mu$ and $\ell\ell$),
lepton+jet ($\ell$j) and two-jet (jj) final states from H/A/h$\ra\tau\tau$ 
are described in detail in
Refs.~\cite{2lepton,hljet,2jets}. In the region of $\rm M _{A} \leq$ 130 
GeV/$c^{2}$ the $\rm gg \rightarrow \rm b \bar{\rm b}h$ cross section is 
comparable to the $\rm gg \rightarrow \rm b \bar{\rm b}A$ production 
cross section, 
therefore it has been included in the analysis.
The relative fractions of these final
states are shown in Table~\ref{table:BR}.  The discovery potential in
the H/A/h$\ra\tau\tau \ra$~lepton+jet channel is re-evaluated
using the cross sections given by the 
routine HQQ \cite{spira_web} and with updated
$\tau$ selection and b-tagging efficiencies. Unlike
Ref.~\cite{hljet}, the recent analysis is extended to large Higgs boson
masses,
and a 5$\sigma$ reach up to m$_{\rm A}\sim$ 650~GeV/$c^2$ at tan$\beta
\sim$~50 is obtained.  The details will be described in an upcoming
note. 

\begin{table}[h]
  
  \caption{Relative fractions of H/A/h$\rightarrow\tau\tau$ final states.}
  \label{table:BR}
  \centering
  \vskip 0.2 in
  \begin{tabular}{|l|c|}
    \hline
     Final state & branching ratio \\
    \hline
     H/A/h$\ra\tau\tau \ra$ e$\mu$+X   & $\sim$6.3\%  \\
     H/A/h$\ra\tau\tau \ra \ell\ell$+X & $\sim$12.5\% \\
     H/A/h$\ra\tau\tau \ra\ell$j+X  & $\sim$45.6\% \\
     H/A/h$\ra\tau\tau \ra$~jj+X & $\sim$41.5\% \\
    \hline
  \end{tabular}
%  \caption{The branching ratios into final states $\tau\tau\ra$X.}
  \vskip 0.1 in
\end{table}

\begin{figure}[t]
  \centering
  \vskip 0.1 in
  \begin{tabular}{cc}
  \begin{minipage}{7.5cm}
    \centering
%    \resizebox{\linewidth}{60 mm}{\includegraphics{\FIG_PATH/effmass_h2tauemu_a200b20_btagging.eps}}
    \resizebox{\linewidth}{60 mm}{\includegraphics{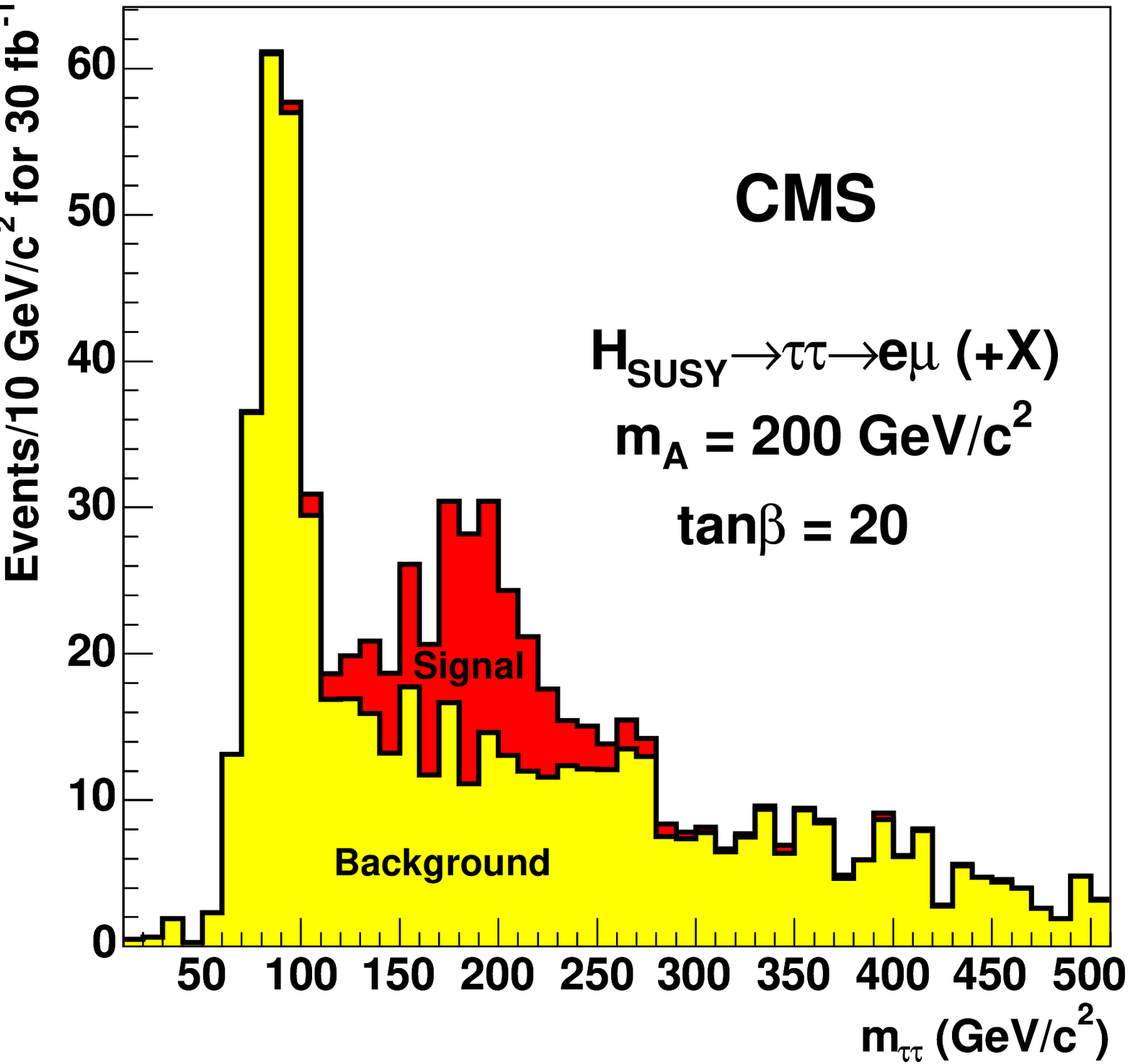}}
    \caption{Reconstructed $\tau\tau$ invariant mass in the e$\mu$ final state
             for the H/A/h~$\rightarrow\tau\tau$ signal (dark) and for the 
             total background (light) with m$_{\rm A}$ = 200 GeV/$c^2$ and 
             tan$\beta$ = 20 for 30~fb$^{-1}$.}
    \label{fig:efmass_emu}
  \end{minipage}
  &
  \begin{minipage}{7.5cm}
    \centering
%    \resizebox{\linewidth}{60 mm}{\includegraphics{\FIG_PATH/effmass_h2tau2l_a200b20_btagging.eps}}
   \resizebox{\linewidth}{60 mm}{\includegraphics{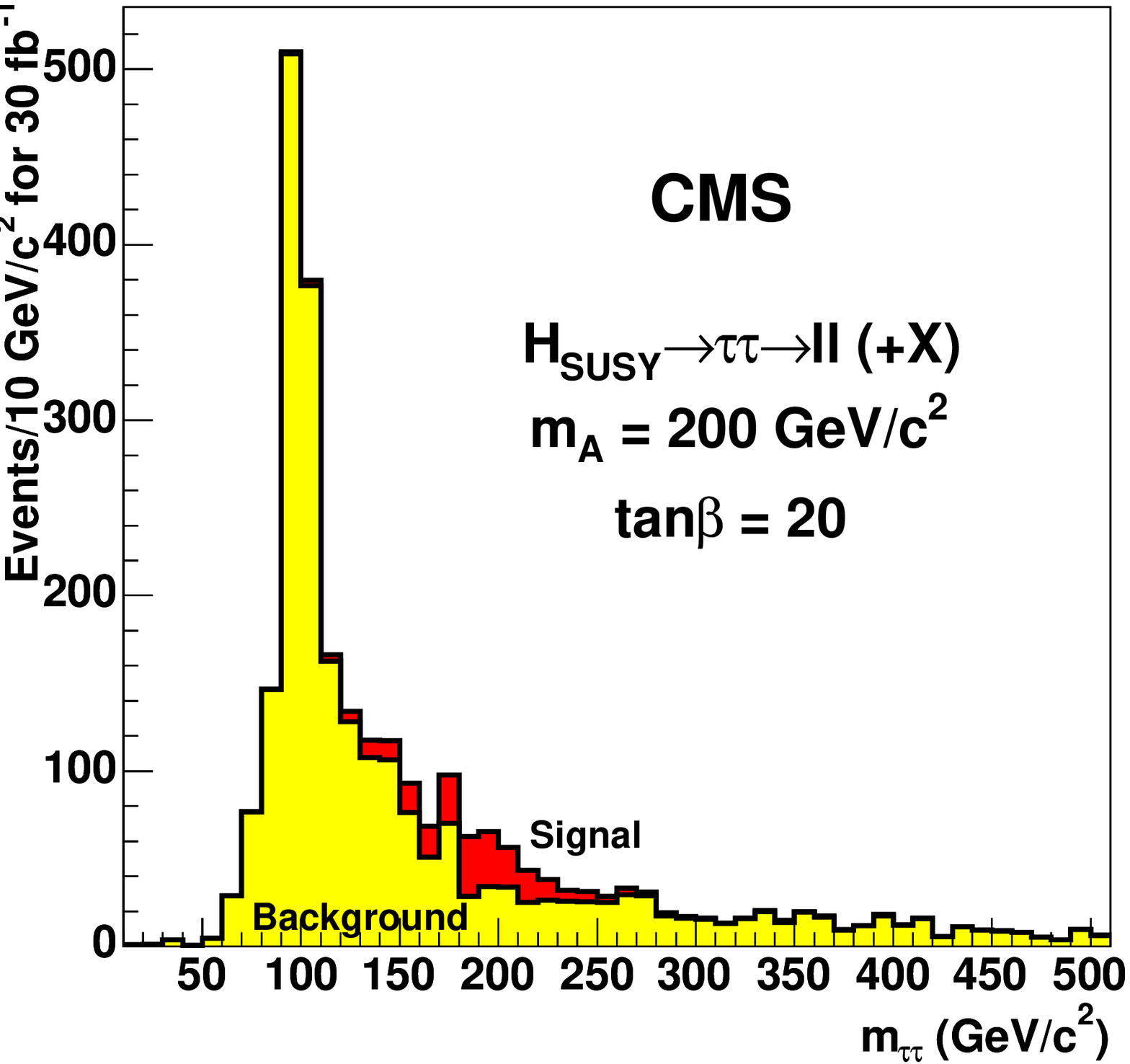}}
    \caption{Reconstructed $\tau\tau$ invariant mass in the $\ell\ell$ final 
             state for the H/A/h~$\rightarrow\tau\tau$ signal (dark) and for 
             the total background (light) with m$_{\rm A}$ = 200 GeV/$c^2$ 
             and tan$\beta$ = 20 for 30~fb$^{-1}$.}
    \label{fig:efmass_ll} 
  \end{minipage}
\\
  \begin{minipage}{7.5cm}
    \centering  
%    \resizebox{\linewidth}{60 mm}{\includegraphics{\FIG_PATH/effmass_a200b20_lj_truestat.eps}}
    \resizebox{\linewidth}{60 mm}{\includegraphics{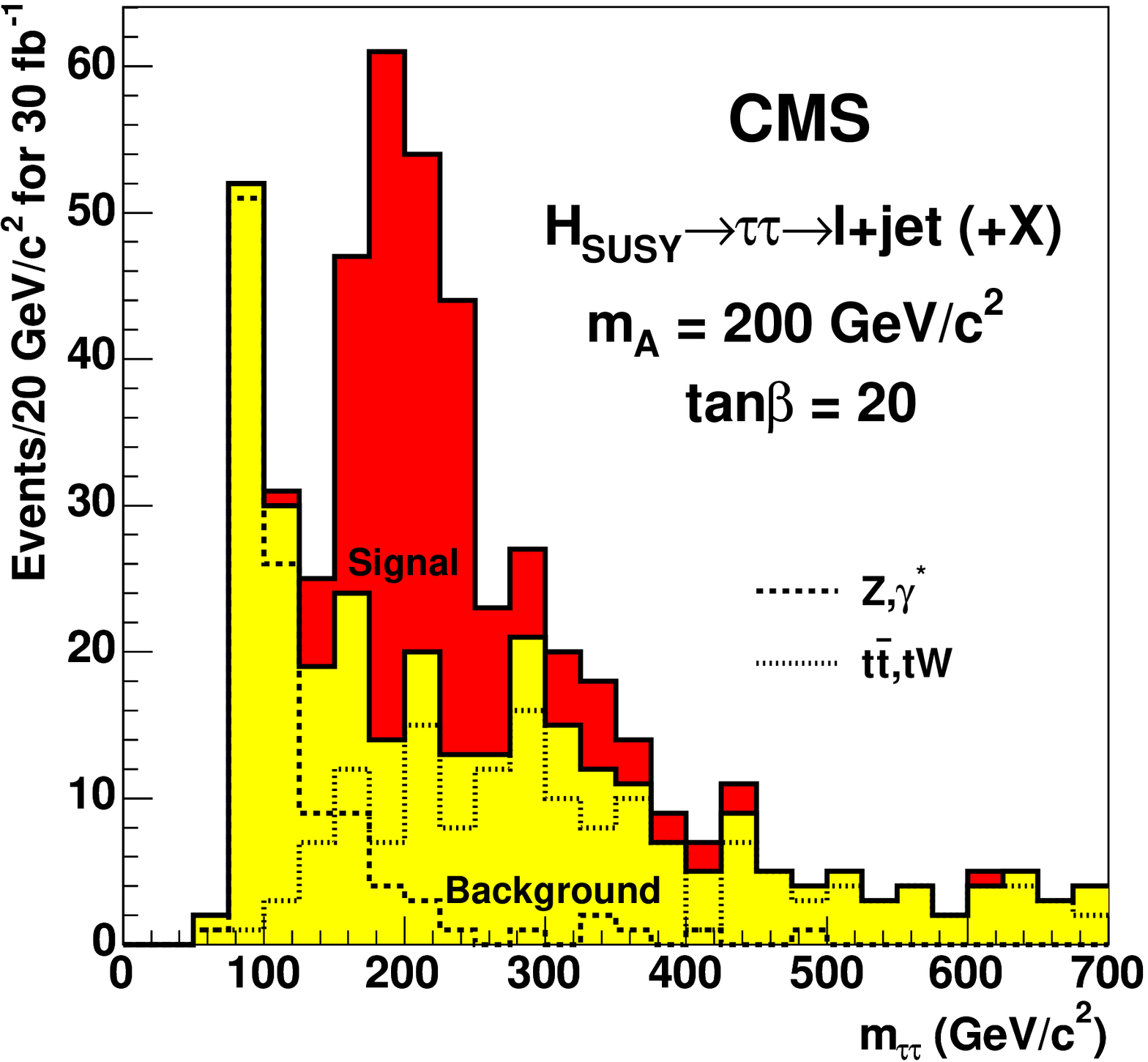}}
    \caption{Reconstructed $\tau\tau$ invariant mass in the lepton+jet final 
             state for the H/A/h~$\rightarrow\tau\tau$ signal (dark) and for 
             the total background (light) with m$_{\rm A}$ = 200 GeV/$c^2$ and 
             tan$\beta$ = 20 for 30~fb$^{-1}$.}
    \label{fig:efmass_lj}
  \end{minipage}
  &
  \begin{minipage}{7.5cm}
    \centering
%    \resizebox{\linewidth}{60 mm}{\includegraphics{\FIG_PATH/ip5_btag_h500_enu_orig.eps}}
    \resizebox{\linewidth}{60 mm}{\includegraphics{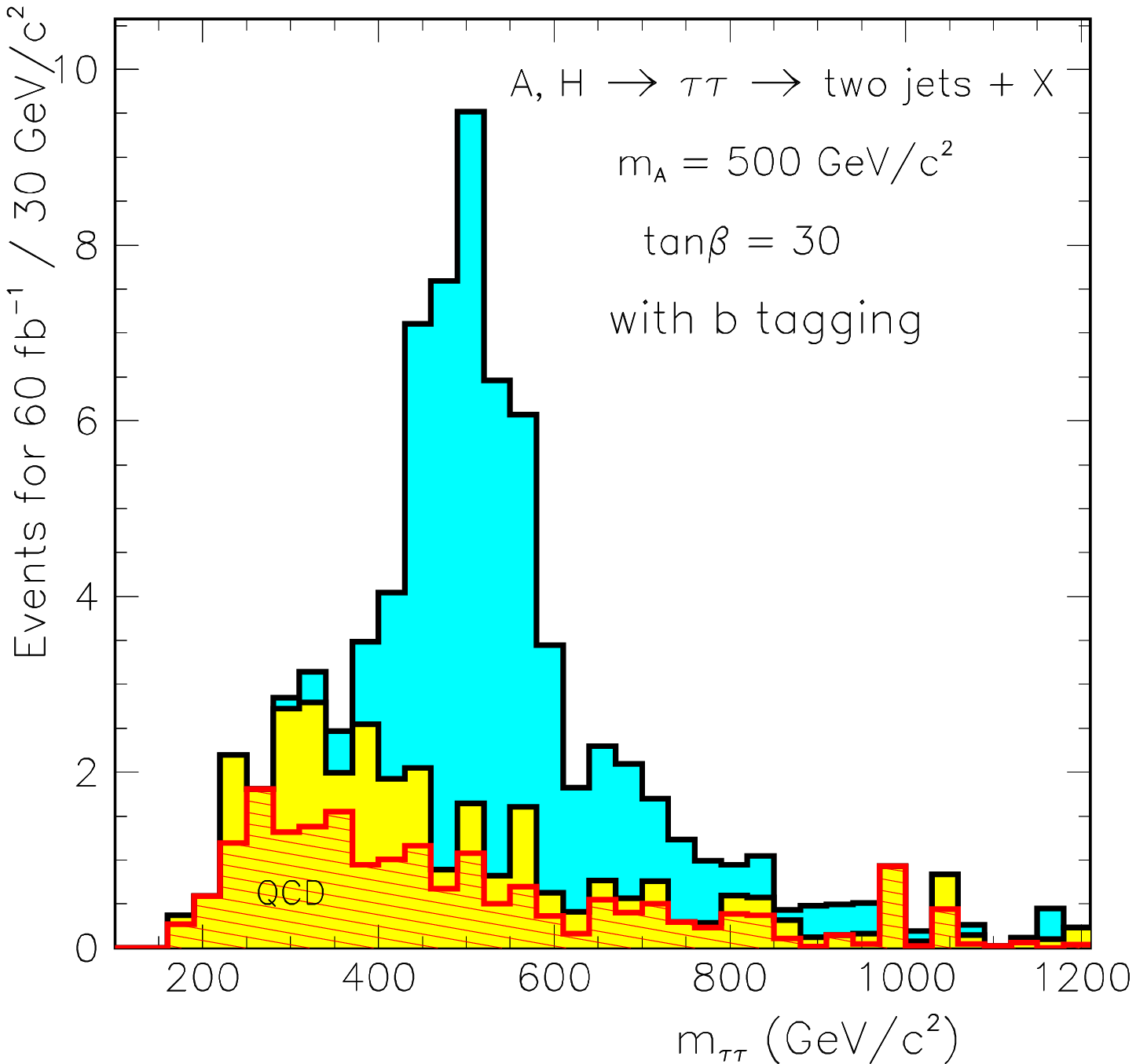}}
    \caption{Reconstructed $\tau\tau$ invariant mass for 
             H/A~$\rightarrow\tau\tau\rightarrow$ 
             2 jets (dark), for the total background
             (light) and for the multi-jet background (dashed) with 
             m$_{\rm A}$ = 500 GeV/$c^2$ and tan$\beta$ = 30 for 60~fb$^{-1}$.}
    \label{fig:efmass_jj}
  \end{minipage}
  \end{tabular}
\end{figure}

 The common backgrounds for all the H/A$\ra\tau\tau$ channels
are the Z,$\gamma^*\rightarrow\tau\tau$ Drell-Yan process, $\rm
t\bar{\rm t}$ production with real and fake $\tau$'s and top
production Wt. The channels with leptons in the final state suffer
from the $\rm b\bar{\rm b}$ background; the W+jet process is
the background for the final states with hadronic $\tau$ decays. For fully
hadronic final states with both $\tau$'s decaying hadronically there is
in addition the QCD multi-jet background with jets faking $\tau$'s, and
for the H/A$\ra\tau\tau \ra \ell\ell$+X channel there is the
additional background from Z,$\gamma^*$ decaying to electron and muon
pairs.

%The signal and backround events were first filtered by triggers
%\cite{DA-HLT-TDR}. 
The hadronic Tau Trigger for the two-jet final state 
was studied with full simulation in Ref. \cite{2jets,DA-HLT-TDR}. For the
e$\mu$, $\ell\ell$ and the $\ell$j final states the trigger was simulated
by selecting the kinematic cuts above the trigger thresholds, and taking
the trigger efficiencies from Ref. \cite{DA-HLT-TDR}. The used triggers
were the Inclusive muon trigger with efficiency
0.9*0.97*0.97 (trigger threshold effect*$\mu$ reconstruction efficiency* 
calorimetric isolation), the Di-electron trigger with efficiency 0.95*0.872*0.946 
per electron (trigger threshold effect*Level-1 e efficiency*Level-2 e efficiency)
and e-$\tau$jet trigger with efficiency 0.95*0.872*0.77*0.95 (e trigger threshold
effect*Level-1 e efficiency*HLT e efficiency*$\tau$ trigger threshold effect).
The backgrounds were
suppressed with lepton tracker isolation, $\tau$ jet identification, $\tau$
tagging with impact parameter, b tagging and jet veto.  The $\tau$ jet
identification \cite{2jets} selects collimated low multiplicity jets
with high p$_{\rm T}$ charged particles.  The hadronic jets are
suppressed by a factor of $\sim$~1000. Tau tagging
\cite{2lepton} exploits the short but measurable lifetime of the $\tau$:
the decay vertex is displaced from the primary vertex. For
b tagging the B hadron lifetime is used to distinguish the associated b
jets from c jets and light quark/gluon jets. The Drell-Yan and 
QCD multi-jet backgrounds are efficiently suppressed with $\rm b$ tagging 
by a factor of
$\sim$~100, but it also suppresses the Higgs boson production with no 
associated b jets. The jet veto is directed against the $\rm t\bar{\rm t}$ 
and $\rm Wt$ backgrounds. The rejection of events with more than one
jet (including the b jet and not counting $\tau$'s) suppresses the $\rm
t\bar{\rm t}$ background by a factor of $\sim$~5 \cite{2jets}.

Despite several neutrinos in all the H/A$\rightarrow\tau\tau$
final states the Higgs boson mass can be reconstructed. The effective
$\tau\tau$ mass was evaluated under the assumption that the neutrinos are 
emitted
along the measured $\tau$ decay products. The projection of the missing 
energy vector onto the neutrinos allowed the neutrino energy estimation.
Uncertainties in the missing energy measurement can lead to negative
neutrino energies.
A significant fraction of the signal events is lost when positive
neutrino energies are required, but the backgrounds from $\rm t\bar{\rm
t}$, Wt and QCD multi-jet events are suppressed, since for these
backgrounds the neutrinos are generally not emitted along the true or
fake $\tau$'s.  The mass resolution can be improved efficiently with a
cut in the $\Delta\phi$ or angular separation between the two $\tau$'s.  For
the two-jet final state in  Ref.~\cite{2jets} a mass reconstruction
efficiency of 53\% has been obtained with the cut $\Delta\phi <$~175$^o$
and with the requirement of one $\rm b$ jet with $\rm E_{\rm T}>$~30~GeV.  
The reconstructed Higgs boson mass is shown in Figs.~\ref{fig:efmass_emu} to
\ref{fig:efmass_jj} for the four $\tau\tau$ final states. The
reconstructed mass peak is a superposition of the H and A signals. In
the region m$_{\rm A}\lsim$ 130 GeV/$c^2$ the contribution from the
lightest Higgs boson h cannot be separated in these channels and is also
included in the signal event rates.

\subsection{Expected discovery reach}

\begin{figure}[h]
\begin{center}
%\mbox{\epsfig{file=\FIG_PATH/discovery_HA_maxmix_30fb.eps,height=9cm,width=12cm}}
\mbox{\epsfig{file=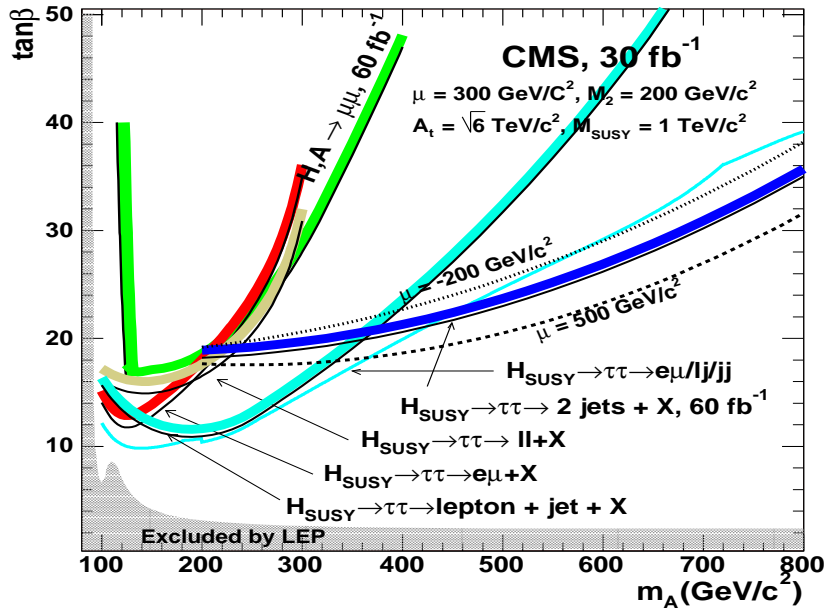,height=9cm,width=12cm}}
\end{center}
\caption{The 5$\sigma$-discovery potential for the heavy neutral MSSM
Higgs bosons as a function of $\rm m_{\rm A}$ and tan$\beta$ with maximal
stop mixing for 30~fb$^{-1}$.  The H/A~$\ra\tau\tau \ra$~two-jet channel 
is shown for 60~fb$^{-1}$.}
\protect\label{fig:discovery_HA}
\end{figure}

The 5$\sigma$-discovery potential for the H/A/h~$\ra\tau\tau$ decay
channels with the e$\mu$, $\ell\ell$ and lepton+jet final states for
30~fb$^{-1}$ and with the two-jet final state for 60~fb$^{-1}$ is shown
in Figure~\ref{fig:discovery_HA}. The 5$\sigma$-discovery reach of the
H/A~$\ra\mu\mu$ decay channel with 60~fb$^{-1}$ is also depicted in the 
figure. The 5$\sigma$-discovery reach obtained after the combination of 
the $\rm e \mu$, lepton+jet 
and two-jet final states from the H/A$\ra \tau\tau$ decay channel is 
shown in Fig.~\ref{fig:discovery_HA} for 30 fb$^{-1}$. The significance
was calculated with the Poisson statistics taking into account the number 
of signal and background events within the Higgs boson mass window.
The combined reach is evaluated
by adding the number of signal and background events from the three
final states in a given  (m$_{\rm A}$, tan$\beta$) point.  This
method, however, can lead to an unsatisfactory result as the analysis of these
final states has been optimized to reach the best possible signal
significance which has led to different background levels. For example
at low values of m$_{\rm A}$ and tan$\beta$ the signal for the
H/A$\ra\tau\tau \ra \ell\ell$+X channel \cite{2lepton} suffers from a
significantly larger Drell-Yan background than that for the
H/A$\ra\tau\tau \ra$~lepton+jet channel. If the $\ell\ell$ final state
is included, the combined reach is smaller than that from the lepton+jet
final state alone. 

\section{Calculation of the cross section measurement uncertainty}

The number of signal events after experimental selection is 
\begin{equation} 
\rm N_{\rm S} = \sigma ^{\rm prod}\times\rm L\times \varepsilon_{\rm sel}, \label{eq:NS1}
\end{equation}
where $\sigma ^{\rm prod}$ is the cross section times branching ratio, 
L is the luminosity and $\varepsilon_{\rm sel}$ is the selection 
efficiency. The measured value of the cross section times branching ratio
is given by
\begin{eqnarray}
  \sigma ^{\rm prod} = \sigma_{0}^{\rm prod} \pm 
  \Delta\rm stat \pm 
  \Delta \rm syst.  
\end{eqnarray}
The considered systematic uncertainties $\Delta \rm syst$ come from the 
luminosity, the experimental selection and the background uncertainties. 
The cross section times branching ratio total uncertainty measurement is 
the quadratic sum of the statistical and systematic uncertainties:

\begin{eqnarray}
\Delta \sigma ^{\rm prod}/ \sigma ^{\rm prod}=  
\sqrt{\rm N_{\rm S} + \rm N_{\rm B}}/ \rm N_{\rm S} \oplus 
\Delta\rm L/\rm L \oplus
\Delta \varepsilon _{\rm sel}/\varepsilon _{\rm sel} \oplus
\Delta \rm N_{\rm B}^{\rm syst}/N_{\rm S},
\end{eqnarray}

where $\rm N_{\rm S}$ and $\rm N_{\rm B}$ are the number of the signal
and background events, $\Delta \rm L$ is the luminosity uncertainty, 
$\Delta \varepsilon _{\rm sel}$ is the experimental selection
uncertainty on the signal and $\Delta \rm N_{\rm B}^{\rm syst}$ is the 
background systematic uncertainty.

\subsection{Statistical and luminosity errors}

The statistical uncertainty from different H/A/h$\rightarrow\tau\tau$ final 
states are combined
using the standard weighted least-squares procedure \cite{ReviewOfPP}.
The measurements are assumed to be uncorrelated and the weighted 
error is calculated as

\begin{equation}
\overline{\sigma ^{\rm prod}} \pm \overline{\Delta\rm stat}
= \frac{\Sigma_i w_i\sigma^{\rm prod}_{i}}{\Sigma_i w_i} \pm (\Sigma_i w_i)^{-1/2}, \label{combined}
\end{equation}

where

\begin{equation}
w_i = 1/(\Delta{\rm stat}_i)^2. \label{weight}
\end{equation}

\begin{figure}[h]
  \centering
  \vskip 0.1 in
    \includegraphics[width=100mm,height=80mm]{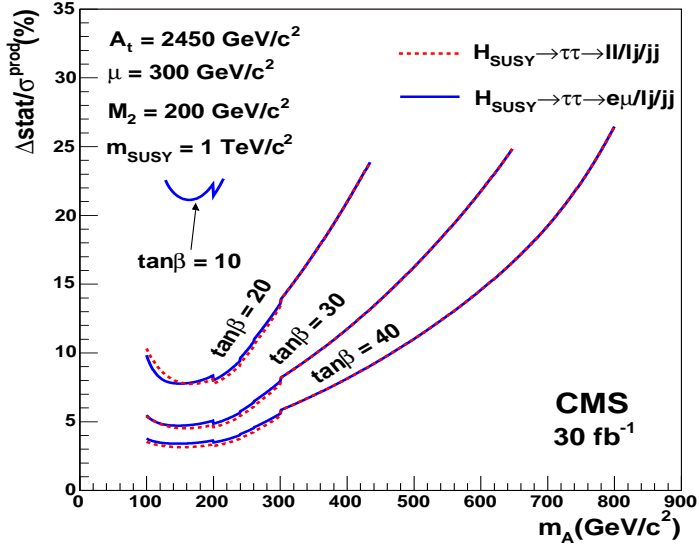}
    \caption{The curves show the statistical uncertainty on the
             cross section times branching ratio measurement as a function
             of m$_{\rm A}$ for tan$\beta$ = 10, 20, 30 and 40 where the 
             statistical significance exceeds 5 $\sigma$.
             The uncertainties correspond to 30 fb$^{-1}$.}
  \label{fig:dtanb_mA_h2tau_emu_lj_jj1}
\end{figure}

Figure \ref{fig:dtanb_mA_h2tau_emu_lj_jj1} shows the statistical uncertainty
on the cross section times branching ratio 
$\Delta\rm stat / \sigma ^{\rm prod}$  for the combined
e$\mu$ + $\ell$j + jj and $\ell\ell$ + $\ell$j + jj
final states as a function of m$_{\rm A}$ for tan$\beta$ = 10, 20, 30 and 40, 
%and for 30 fb$^{-1}$. 
where the signal significance for 30 fb$^{-1}$ exceeds 5$\sigma$.
The drop in the curves at m$_{\rm
A}$ = 300 GeV/$c^2$ is due to fully leptonic final states (e$\mu$ and
$\ell\ell$) being accessible and included in the tan$\beta$ measurement
in the region from m$_{\rm A}$ = 100 GeV/$c^2$ to m$_{\rm A}$ = 300
GeV/$c^2$.  Similarly, a small decrease is visible at m$_{\rm A}$ = 200
GeV/$c^2$ due to the fully hadronic final state being analyzed only in
the region from m$_{\rm A}$ = 200 GeV/$c^2$ to m$_{\rm A}$ = 800
GeV/$c^2$.  The statistical uncertainty ranges from 3\% to 25\% depending on
tan$\beta$ and $\rm m_{\rm A}$. 

The uncertainty on the luminosity measurement is assumed to be 5\%.

\subsection{Uncertainty on the signal selection}

The uncertainty on the signal selection efficiency related to the absolute
calorimeter energy scale, $\tau$ and b-tagging efficiency was taken into 
account as described in the following.

The calorimeter energy scale introduces an error on the selection efficiency 
since jets and missing $\rm E_{T}$ threshold are required.
The full simulation of two tau-jet signal events shows that the assumption of 
1 \% uncertainty on the calorimeter energy scale \cite{DA-HLT-TDR} leads to 
$\simeq$ 2.9 \% uncertainty on the signal selection efficiency.

The single $\rm b$-tagging efficiency was evaluated in the following way.
Two samples of semileptonic $\rm t \bar{\rm t}$ events selected for the
top quark mass measurement with double $\rm b$ tagging and single $\rm b$ 
tagging respectively were used. The single $\rm b$-tagging efficiency 
is obtained as the ratio of double vs single $\rm b$-tagged events in the
samples, with an additional top mass constraint. Since about 10000 double 
$\rm b$-tagged events were obtained after all selections with 30 fb$^{-1}$ 
\cite{lars}, the statistical error of the $\rm b$-tagging efficiency was
$\simeq$ 1\%. Nevertheless, the b-jet purity and the background 
contribution for the single b-tagged semileptonic $\rm t \bar{\rm t}$ 
events still have to be investigated. A value of 2 \% for the single 
b-tagging efficiency uncertainty is taken in this study.

The uncertainty on the $\tau$ tagging can be evaluated from the measured 
ratio of
$\rm Z \rightarrow \tau \tau \rightarrow \mu + \tau ~ \rm jet$ and 
$\rm Z \rightarrow \mu \mu$ events. The background contribution to 
$\mu$+$\tau$ jet final state from $\rm t \bar{\rm t}$ and W+jet events 
is reduced with the cut on muon plus missing 
$\rm E_{\rm T}$ transverse mass , $\rm m(\ell, \rm E_{\rm T}^{\rm miss})$,  
and the Z mass constraint. About 10000  
$\rm Z \rightarrow \tau \tau \rightarrow \mu+ \tau$ jet 
events are expected 
with 30 fb$^{-1}$ after muon and $\tau$-jet selections and a cut on
$\rm m(\ell, \rm E_{\rm T}^{\rm miss})<$ 30 GeV/$c^2$. 
Thus, the statistical uncertainty on the $\tau$-tagging efficiency is about 
1\%  ($\rm Z \rightarrow \mu \mu$ statistical error is negligible). The cut
on $\rm m(\ell, \rm E_{\rm T}^{\rm miss})$ introduces however about 1 \%
uncertainty due to the calorimeter energy scale uncertainty. The remaining
uncertainty may come from the evaluation of $\rm t \bar{\rm t}$ and W+jet
background under the Z mass peak (about 2500 background events are 
expected before the Z mass constraint with 30 fb$^{-1}$) and the uncertainty
on the muon reconstruction efficiency. Due to these reasons another 2 \% 
was added as a conservative, preliminary estimate leading thus to the 
total uncertainty on the $\tau$-jet identification efficiency of the 
order of 2.5 \%.

Therefore, the signal experimental selection uncertainty due to the 
calorimeter energy scale, $\tau$ and b-jet tagging efficiency is 4.3 \%. 

\subsection{Background uncertainty}

The background uncertainty was evaluated for one representative point of the 
(m$_{\rm A},\tan\beta$) parameter space m$_{\rm A}$ = 300 GeV/$c^2$, 
$\tan\beta$ = 20 and for the H/A$\ra\tau\tau \ra$~lepton+jet channel. 
The point in the parameter space was chosen close to the 5$\sigma$ limit where 
the number of signal events is lowest, and therefore the 
error $\Delta \rm N_{\rm B}/N_{\rm S}$ is largest. 

The background after all cuts within a chosen mass window was estimated by 
fitting the m$_{\tau\tau}$ distribution of the expected background 
and signal distributions. The error of the fit gives the background 
uncertainty. The signal mass distribution was approximated with a 
Gaussian function. The background, however, consists of several different 
components 
with different mass distribution shapes. The Drell-Yan background 
forms a high peak around the Z mass with a long high mass tail. 
The $\rm t\bar{\rm t}$ and single top backgrounds form a much wider mass 
distribution. The W+j and $\rm b \bar{\rm b}$ backgrounds were not taken 
into account in the fit since their contribution to the total 
background is negligible ($\leq$ 5\%).

The shape of the background mass distributions can be obtained from Monte Carlo
or directly from the data by relaxing the cuts chosen to suppress the 
background. 
The top background can be enhanced by double b tagging (relaxing the central
jet veto) which suppresses the signal and Drell-Yan background. The shape 
of the $\rm t \bar{\rm t}$ background is best fitted with a Landau 
distribution. The small change of the shape due to the different event 
selections was neglected.  
The Drell-Yan background shape could be evaluated with the suppression of 
the b tagging requirement.
In that case the signal was very small compared to the background. 
The shape of the Drell-Yan background was also best fitted with a 
Landau distribution. 

Using the shapes for the signal and background distributions, the mass 
distribution can be fitted varying the number of signal, Drell-Yan and top 
events. The fit parameters were then used to evaluate the number of 
background events and its uncertainty in the signal mass window.  
For the chosen point m$_{\rm A}$ = 300 GeV/$c^2$ and $\tan\beta$ = 20 this 
procedure gives 80.5 signal events and 25.9 background events with 
$\Delta\rm N_{\rm B}$ = 8.1, which corresponds to a background 
uncertainty $\Delta \rm N_{\rm B}/N_{\rm S}$ = 10\%. This uncertainty is used 
then as a conservative estimate for the whole (m$_{\rm A},\tan\beta$) 
parameter space and for $\ell \ell$ and two $\tau$-jet final states as well.

\subsection{Total systematic uncertainty}

For the chosen point, m$_{\rm A}$ = 300 GeV/$c^2$ and $\tan\beta$ = 20, the 
total systematic uncertainty is 12 \% (the dominat contribution comes from 
the background uncertainty 10 \%) and is of the same order as the statistical 
uncertainty. The background uncertainty decreases for higher values 
of $\tan\beta$ at a given value of m$_{\rm A}$.

\section{Calculation of the tan$\beta$ measurement uncertainty}

It is clear that the determination of tan$\beta$ alone without any knowledge 
of the other SUSY parameters is not possible, since the branching ratios
develop a significant dependence on $\mu$, M$_{2}$, A$_{\rm t}$ and 
M$_{\rm SUSY}$.
For example, the gaugino sector is sensitive to the values of $\mu$ and
M$_{2}$, and the stop sector and stop mixing is sensitive to the values of
$\mu$ and A$_{\rm t}$. Nevertheless, in order to illustrate the potential of a
cross section times branching ratio measurement in the tan$\beta$ 
determination, the SUSY parameters were fixed to the chosen SUSY scenario, 
but without taking into account the uncertainties, since they are still 
unknown. Nonetheless, to give an idea of the sensitivity of the signal 
rate to the 
SUSY parameters, those were varied by 20 \% around the nominal values, as it 
will be discussed in Section 4.3.

Ignoring in this approach the uncertainties of the other SUSY parameters,
the accuracy of the tan$\beta$ measurement is due to the uncertainty on the 
cross section measurement $\Delta \sigma ^{\rm prod}$ and the theoretical
uncertanty of the cross section calculation.

\begin{figure}[h]
  \centering
  \vskip 0.1 in
  \includegraphics[height=70mm,width=80mm]{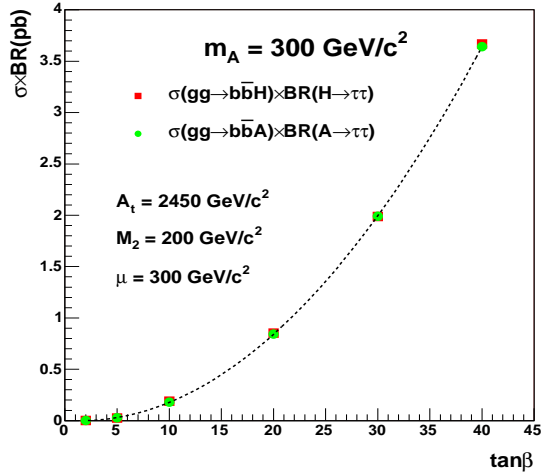}
      \caption{Cross section times branching ratio for
               $\rm gg\rightarrow b\bar{b}H/A, H/A\rightarrow\tau\tau$
               calculated with the programs of Ref.~\cite{spira_web}. }
  \label{fig:bbh_crosssection}
\end{figure}

The branching ratio BR(H/A~$\rightarrow\tau\tau$) is approximately constant
at large tan$\beta$. At large tan$\beta$ the total decay width is
dominated by Higgs boson decay to heavy down type fermions, $\tau^+
\tau^-$ and $\rm b\bar{\rm b}$ pairs, for which the decay widths have similar
tan$\beta$ dependence. If the SUSY corrections, which are different for
the bottom and $\tau$ Yukawa couplings, are not large, the tan$\beta$
dependence cancels out in the ratio
$\Gamma$(H/A~$\rightarrow\tau\tau$)/$\Gamma_{\rm tot}$, which becomes
approximately constant. The counting of signal events measures the 
total rate $\sigma\times$BR into the chosen final state, which is therefore
approximately proportional to tan$^2\beta$. The total rate for
the neutral Higgs boson production as a function of tan$\beta$
is shown in Fig. \ref{fig:bbh_crosssection} for 
m$_{\rm A}$ = 300 GeV/$c^{2}$.

At large tan$\beta$ the production rate can be written as 
\begin{equation}
\sigma ^{\rm prod} = \rm tan^2\beta \times\rm X,
\end{equation}
where X is the tan$\beta$ independent part of the production rate. 
The measured value of tan$\beta$ is given by
\begin{eqnarray}
  \tan\beta = \tan\beta_0 \pm 
  \frac{1}{2}\Delta\sigma ^{\rm prod} \pm 
  \frac{1}{2}\Delta \rm X  
\end{eqnarray}
where $\Delta \rm X$ consist of the theoretical uncertainties of
the production cross section and the branching ratio, and the 
uncertainty on the cross section due to the uncertainty on the
Higgs boson mass measurement. The total error of $\rm tan \beta$ is 
taken as the quadratic sum of $\Delta\sigma ^{\rm prod}$ and $\Delta \rm X$:
\begin{eqnarray}
  \Delta \rm tan\beta / tan\beta =  
\frac{1}{2}\Delta\sigma ^{\rm prod}/\sigma ^{\rm prod} \oplus
\frac{1}{2}\Delta\rm X/\rm X.
\end{eqnarray}

\subsection{Theoretical uncertainty}

The uncertainty on the Next-to-Leading Order 
cross sections for the gg~$\ra \rm b \overline{\rm b}$H/A/h 
process has been shown to be 20$-$30\% for the
total rate \cite{hep-ph/0309204,plehn}. It depends, however, on the
transverse momentum range of the spectator b quarks and reduces to
10$-$15\% with the requirement of p$_{\rm T}^{\rm b,\bar{\rm b}}\gsim$ 20
GeV/$c$ \cite{hep-ph/0309204,hep-ph/0204093,dawson}. Therefore,
the question about requiring two b jets per event with jet E$_{\rm T}>$
20~GeV arises naturally.  Table \ref{table:2btagging} shows the number
of signal and background events for the H/A~$\ra\tau\tau
\ra$~lepton+jet+X channel with m$_{\rm A}$ = 200 GeV/$c^2$ and
tan$\beta$ = 20 for one $\rm b$-tagged jet in the event (plus a veto on
additional jets) and for events with two $\rm b$-tagged jets. Although the
theoretical uncertainty is smaller for events with two $\rm b$-tagged 
jets with jet E$_{\rm T}>$ 20 GeV, 
the decrease of the signal statistics due to low 
reconstruction and b-tagging efficiency of soft b jets \cite{CMSNote2001/019}
increases the measurement uncertainty. Therefore, only one b jet
per event should be tagged in this study.  The theoretical
uncertainty of about 20\% is adopted for the production cross section 
according to
Refs.~\citer{hep-ph/0309204,dawson}. For the branching ratio the uncertainty 
is 3\% and it is related to the uncertainties of the SM input parameters.

\begin{table}[h]
 
  \caption{The uncertainty on the tan$\beta$ measurement for the
           H/A~$\ra\tau\tau \ra$~lepton+jet+X channel for 30 fb$^{-1}$
           with one or two b-tagged jets with jet E$_{\rm T}>$ 20 GeV.}
  \label{table:2btagging}
 \vskip 0.1 in 
  \begin{center}
  \begin{tabular}{|l|c|c|c|c|c|c|}
   \hline
    m$_{\rm A}$ = 200 GeV/$c^2$, tan$\beta$ = 20 & N$_{\rm S}$ & N$_{\rm B}$ & signif.
    & $\sqrt{\rm N_{\rm S} + N_{\rm B}}/\rm N_{\rm S}$ & $\Delta\sigma/\sigma$ & $\Delta\rm tan\beta/\rm tan\beta$$^*$ \\
   \hline
%    1b-tagging+jet veto & 205 & 91  & 21.5$\sigma$ & 8.4\%  & 20\% & 16.7\%      \\
%    2b-tagging          & 12  & 57  & 1.6$\sigma$  & 69.2\% & 10-15\%    & 42.1-44.6\% \\
    1b-tagging+jet veto & 157 & 70  & 18.8$\sigma$ & 9.6\%  & 20\%    & 17.3\%      \\
    2b-tagging          & 9   & 44  & 1.3$\sigma$  & 80.9\% & 10-15\% & 48.0-50.5\% \\
   \hline
\multicolumn{7}{l}{ \small $^{*)}$ Statistical + theoretical cross section 
uncertainties only}
  \end{tabular}
  \end{center}
\end{table}

\subsection{Mass measurement uncertainty}

\begin{figure}[h]
  \centering
  \vskip 0.1 in
  \begin{tabular}{ccc}
  \begin{minipage}{5cm}
    \centering
  \includegraphics[height=70mm,width=50mm]{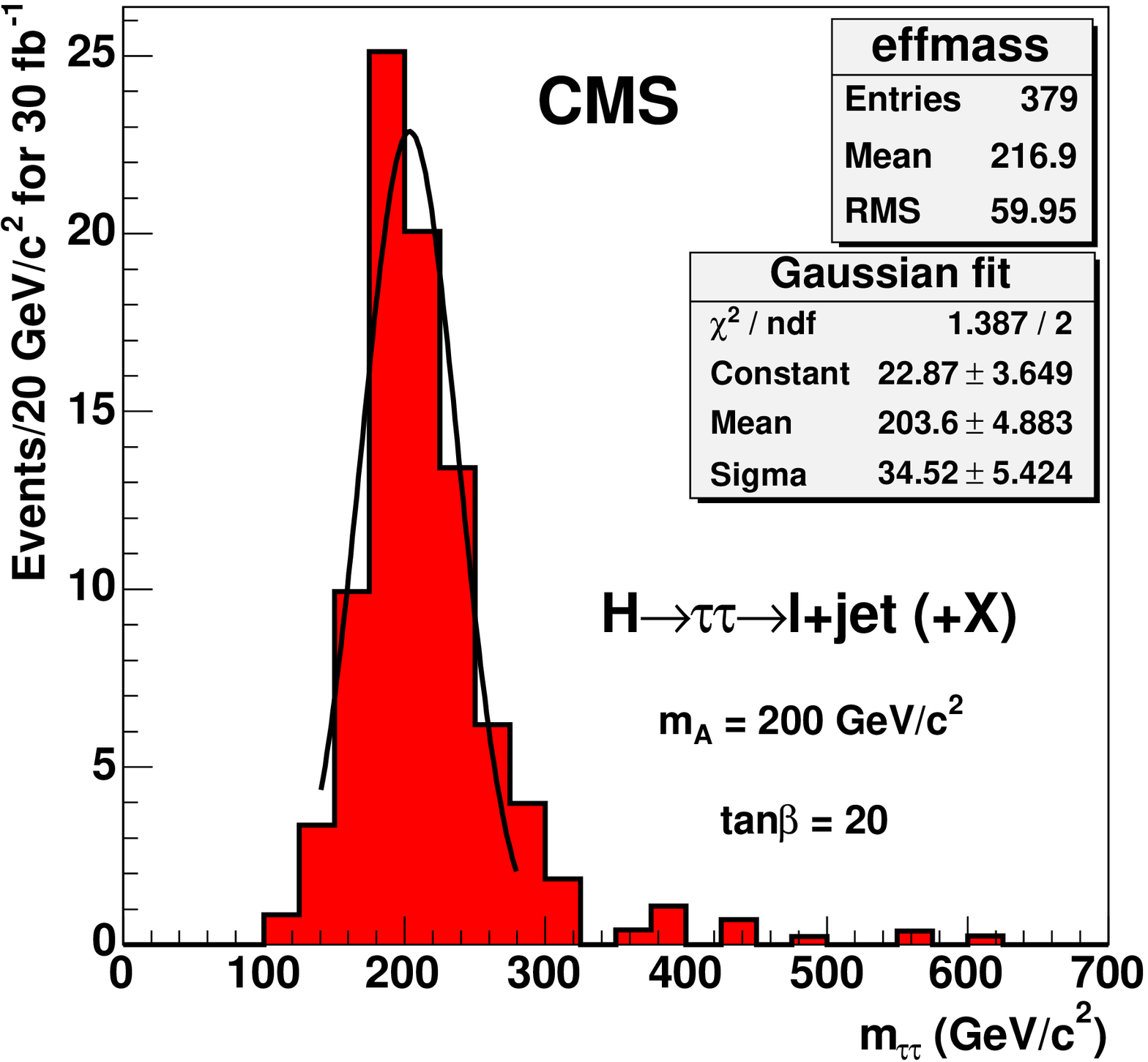}
      \caption{The reconstructed mass of H with Gaussian fit.}
  \label{fig:fit_H}
  \end{minipage}
  &
  \begin{minipage}{5cm}
    \centering
   \includegraphics[height=70mm,width=50mm]{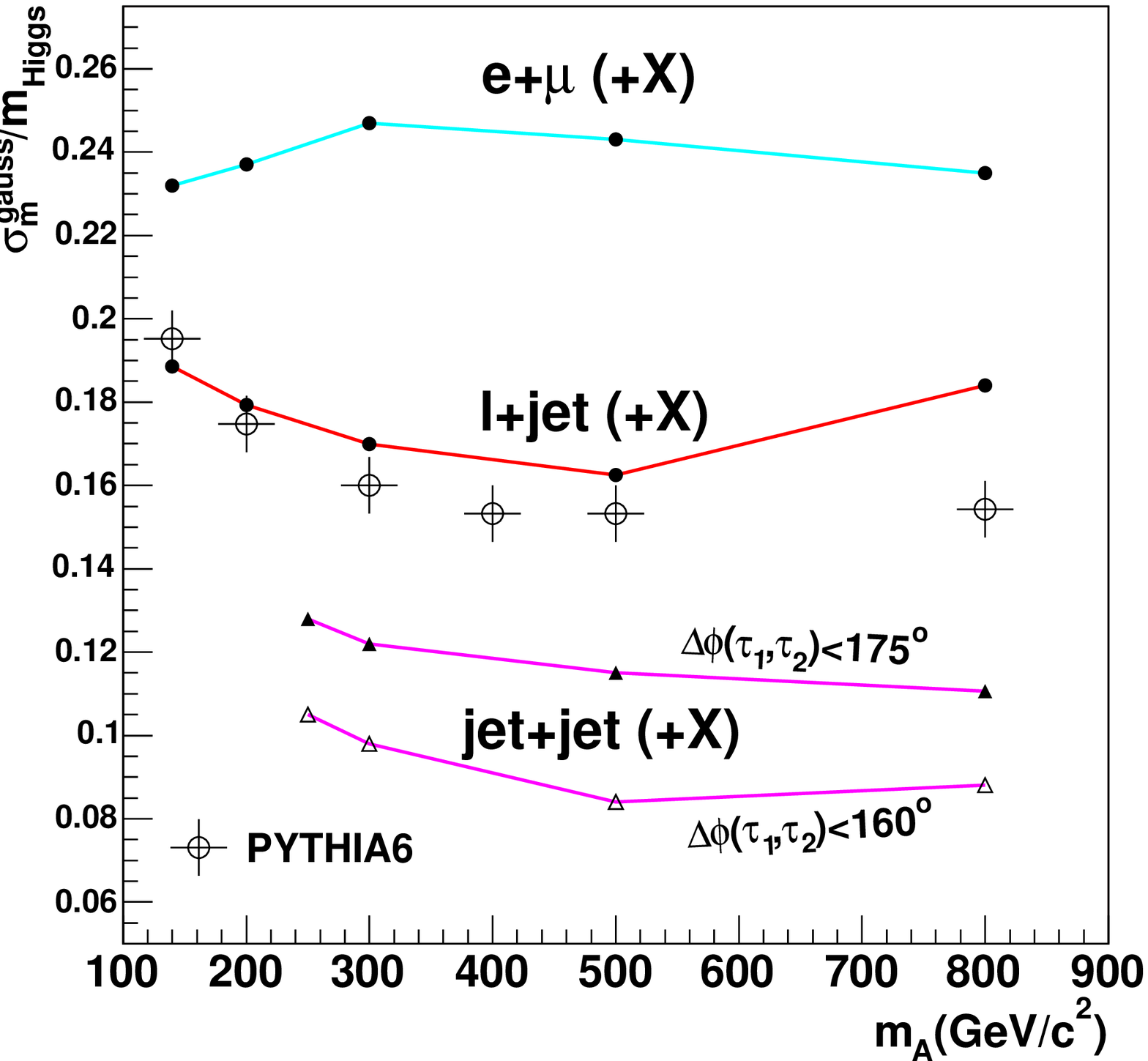}
      \caption{The mass resolution as a function of m$_{\rm A}$ \cite{2lepton}.}
  \label{fig:mreso}
  \end{minipage}
  &
  \begin{minipage}{5cm}
    \centering
  \includegraphics[height=70mm,width=50mm]{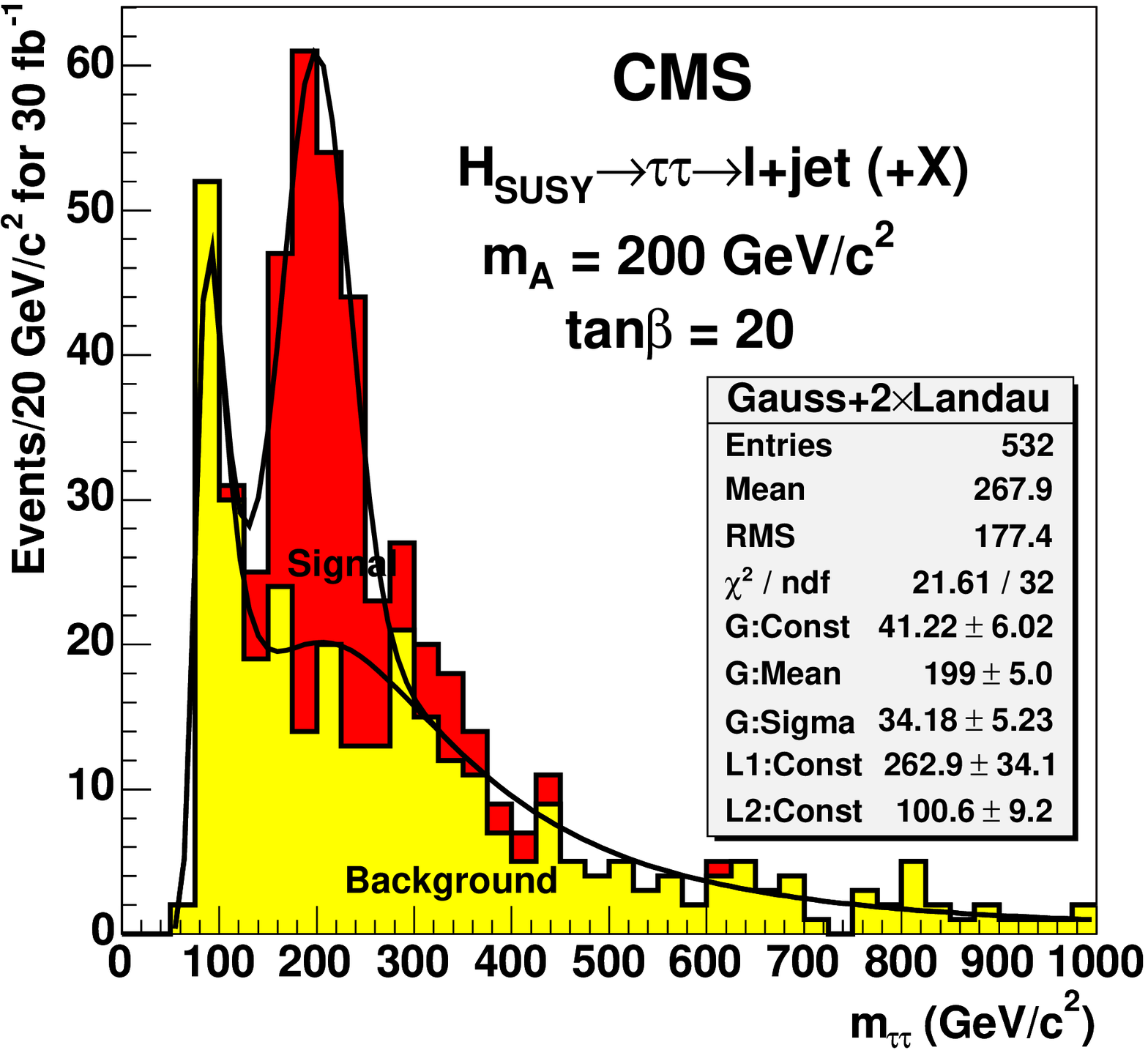}
      \caption{Fit of the signal+background mass distribution.}
  \label{fig:fit_SB}
  \end{minipage}
  \end{tabular}
\end{figure}

Since the value of the cross section depends on the Higgs boson mass, the 
uncertainty on the mass measurement leads to an uncertainty on the signal
rate. The Higgs mass is measured using the different final states,
and the cross section uncertainties due to the mass measurement uncertainty
are combined using equations \ref{combined} and \ref{weight}
which give smallest weight to the channels with largest uncertainty. 
The mass resolution is estimated by fitting the reconstructed m$_{\tau \tau}$ 
distribution after all selections with a
Gaussian function (Fig. \ref{fig:fit_H}). The resolution 
is almost constant as a function of m$_{\rm A}$, $\sim$ 24\%
for the leptonic final state, $\sim$ 17\% for the lepton+jet final state
and $\sim$ 12\% for the hadronic final 
state as shown in Fig. \ref{fig:mreso}. 
The fit of the signal plus background distributions (Fig. \ref{fig:fit_SB}) 
gives the same Higgs boson mass resolution as the fit of the signal only. The 
uncertainty on the mass measurement is calculated from the Gaussian fit of the
mass peak as $\sigma_{\rm Gauss}/\sqrt{\rm N_{\rm S}}$, and the
error induced to the cross section ($\Delta\sigma(\Delta\rm m)$) 
is estimated by varying the cross section 
for Higgs masses
m$_0$ and m$_0\pm\sigma_{\rm Gauss}/\sqrt{\rm N_{\rm S}}$. At 5$\sigma$ limit 
where the signal statistics is lowest, the uncertainty on the mass measurement
brings 5 - 6\% uncertainty on the tan$\beta$ measurement.

\subsection{Dependence on the SUSY parameters}

To give an estimation of the sensitivity 
of the signal rate to the SUSY parameters, those were varied 
by 20\% around the nominal values. 
The variation of the signal rate within the expected 
discovery region is at most $\sim$11\% which leads to a maximum of 6\% 
uncertainty on the tan$\beta$ measurement. At low m$_{\rm A}$ the uncertainty 
is even smaller. As already stated, in reality, it is expected that the cross 
section measurement will be entered in a global fit together with the other 
relevant measurements to determine all SUSY parameters simultaneously.

\section{Measurement of tan$\beta$} \label{sec:Results}

\subsection{H/A $\ra\tau\tau$}

Table \ref{table:errors} shows the statistical uncertainty on the
tan$\beta$ measurement and the uncertainty on the mass measurement 
for each individual final state and for the
combined final states from H/A~$\ra \tau\tau$ for 30~fb$^{-1}$.  
The total estimated uncertainty is shown for the combined final states.
As described in the previous section it includes the 
following uncertainties: statistical, theoretical, luminosity, 
mass measurement, uncertainty on the experimental signal 
selection efficiency and the background uncertainty. 
The results are shown for
the region of the (m$_{\rm A}$, tan$\beta$) parameter space where the
statistical significance exceeds 5$\sigma$. Close to the 5$\sigma$ limit
the statistical uncertainty is of the order of 11 - 12\%, but it
decreases rapidly for increasing tan$\beta$.

\begin{table}[h]
\centering
%\vskip 0.1 in
\caption{Statistical uncertainty on the tan$\beta$ measurement and the
         uncertainties due to the mass measurement for individual
         and combined final states in
         four (m$_{\rm A}$,tan$\beta$) parameter
         space points for 30 fb$^{-1}$ with better than 5$\sigma$ 
         significance. 
         The total uncertainty $\Delta \rm tan\beta / tan\beta$ in the Table 
         includes the following uncertainties: statistics, mass measurement, 
         cross section and branching ratio theoretical uncertainties, 
         luminosity, experimental selection efficiency and background 
         uncertainties.}
\label{table:errors}
 \vskip 0.2 in
\begin{tabular}{|c||c|c|c|c|c|c|c|c|}
\hline
\multirow{2}{2cm}{\centering \Large 30 fb$^{-1}$}
%& \multicolumn{4}{|c|}{Stat.+lumi.+theor. error\% (stat. error\%)} \\
%%%%& \multicolumn{4}{|c|}{$\Delta \rm tan\beta / tan\beta$ \% ($\Delta$stat \%)} \\
%%%% \cline{2-5}
& \multicolumn{2}{|c|}{\begin{minipage}{2.5cm}
m$_{\rm A}$~=~200~GeV/$c^2$ \\
tan$\beta$ = 20
\end{minipage}}
& \multicolumn{2}{|c|}{\begin{minipage}{2.5cm}
m$_{\rm A}$~=~200~GeV/$c^2$ \\
tan$\beta$ = 30
\end{minipage}}
& \multicolumn{2}{|c|}{\begin{minipage}{2.5cm}
m$_{\rm A}$~=~500~GeV/$c^2$ \\
tan$\beta$ = 30
\end{minipage}}
& \multicolumn{2}{|c|}{\begin{minipage}{2.5cm}
m$_{\rm A}$~=~500~GeV/$c^2$ \\
tan$\beta$ = 40
\end{minipage}}\\
\cline{2-9}
 & $\Delta$stat & $\Delta\sigma(\Delta\rm m)$ & $\Delta$stat & $\Delta\sigma(\Delta\rm m)$
 & $\Delta$stat & $\Delta\sigma(\Delta\rm m)$ & $\Delta$stat & $\Delta\sigma(\Delta\rm m)$ \\
\hline
H/A$\rightarrow\tau\tau\rightarrow$e$\mu$   & 8.95\% & 4.82\% & 4.85\% & 3.27\% & - & - & - & -  \\
H/A$\rightarrow\tau\tau\rightarrow\ell\ell$ & 7.96\% & 3.50\% & 4.08\% & 2.37\% & - & - & - & -  \\
H/A$\rightarrow\tau\tau\rightarrow\ell$j    & 4.81\% & 2.46\% & 2.84\% & 1.65\% & - & - & 8.40\% & 4.82\% \\
H/A$\rightarrow\tau\tau\rightarrow$jj       & 13.7\% & 4.73\% & 8.25\% & 3.21\% & 12.4\% & 5.82\% & 8.45\% & 4.44\% \\
\hline
\hline
\multirow{2}{3cm}{\begin{minipage}{2.cm}\bf Combined \\ e$\mu$+$\ell$j+jj\end{minipage}}
& 4.05\% & 1.99\% & 2.35\% & 1.34\% & 9.09\% & 4.28\% & 5.96\% & 3.26\% \\
\cline{2-9}
\cline{2-9}
& \multicolumn{2}{|c|}{$\Delta \rm tan\beta / tan\beta$  }
& \multicolumn{2}{|c|}{$\Delta \rm tan\beta / tan\beta$  }
& \multicolumn{2}{|c|}{$\Delta \rm tan\beta / tan\beta$  }
& \multicolumn{2}{|c|}{$\Delta \rm tan\beta / tan\beta$  } \\
\cline{2-9}
& \multicolumn{2}{|c|}{ 12.59\% }
& \multicolumn{2}{|c|}{ 12.06\% }
& \multicolumn{2}{|c|}{ 15.46\% }
& \multicolumn{2}{|c|}{ 13.57\% } \\
\hline
\multirow{2}{3cm}{\begin{minipage}{2.cm}\bf Combined \\ $\ell\ell$+$\ell$j+jj\end{minipage}}
& 3.94\% & 1.85\% & 2.24\% & 1.25\% & 9.09\% & 4.28\% & 5.96\% & 3.26\% \\
\cline{2-9}
\cline{2-9}
& \multicolumn{2}{|c|}{$\Delta \rm tan\beta / tan\beta$  }
& \multicolumn{2}{|c|}{$\Delta \rm tan\beta / tan\beta$  }
& \multicolumn{2}{|c|}{$\Delta \rm tan\beta / tan\beta$  }
& \multicolumn{2}{|c|}{$\Delta \rm tan\beta / tan\beta$  } \\
\cline{2-9}
& \multicolumn{2}{|c|}{ 12.53\% }
& \multicolumn{2}{|c|}{ 12.03\% }
& \multicolumn{2}{|c|}{ 15.46\% }
& \multicolumn{2}{|c|}{ 13.57\% } \\
\hline
%\multicolumn{5}{l}{ \small $^{*)}$ Significance $< 5\sigma$ }
\end{tabular}
\vskip 0.1 in
\end{table}

As shown in the table, the highest statistical accuracy of about 5\% for
m$_{\rm A}$ = 200 GeV/$c^2$ and tan$\beta$ = 20, is obtained with the
lepton+jet final state. Combining other channels with the lepton+jet
channel in this mass range improves the precision only slightly. 
%For m$_{\rm A}$ = 500 GeV/$c^2$ and tan$\beta$ = 30, the lepton+jet channel
%still yields higher statistical accuracy, about 10\%, than the two-jet
%channel ($\sim$~12\%). 
The difference between the fully leptonic
channels (e$\mu$ and $\ell\ell$) is small: the statistical uncertainty
is slightly smaller for the e$\mu$ channel, if m$_{\rm A}$ is close to
the Z peak, but already at m$_{\rm A}$ = 200 GeV/$c^2$ the final state
with any two leptons yields better statistics and lower uncertainties.
The combined $\ell\ell$+$\ell$j+jj channel yields a slightly smaller
statistical error at tan$\beta$ = 20 than the combined
e$\mu$+$\ell$j+jj channel despite the larger backgrounds in the
$\ell\ell$ final state.

\begin{figure}[h]
  \centering
  \vskip 0.1 in
    \includegraphics[width=100mm,height=80mm]{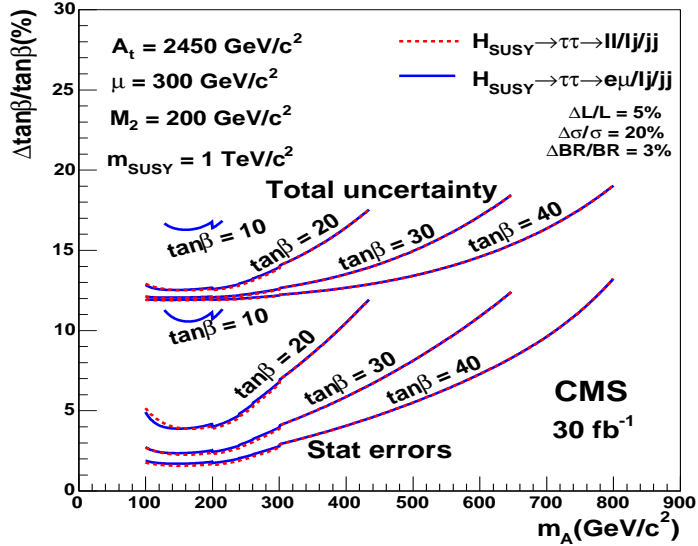}
    \caption{Measurement uncertainty on tan$\beta$: the three lower curves 
             show the uncertainty on tan$\beta$ measurement when only the 
             statistical uncertainty is taken into account. The three upper 
             curves show the total uncertainty evaluated as described in the 
             text. The uncertainty correspond to 30 fb$^{-1}$.}
  \label{fig:dtanb_mA_h2tau_emu_lj_jj}
\end{figure}

Figure \ref{fig:dtanb_mA_h2tau_emu_lj_jj} shows the statistical
uncertainty and the
statistical plus systematic uncertainties on tan$\beta$ for the combined
e$\mu$ + $\ell$j + jj and $\ell\ell$ + $\ell$j + jj final states as a 
function of m$_{\rm A}$ for tan$\beta$ = 20, 30 and 40, and for 30 fb$^{-1}$. 
The total uncertainty ranges from 12\% to 19\% depending on
tan$\beta$ and $\rm M_{\rm A}$. 

\begin{figure}[p]
  \centering
  \includegraphics[width=110mm]{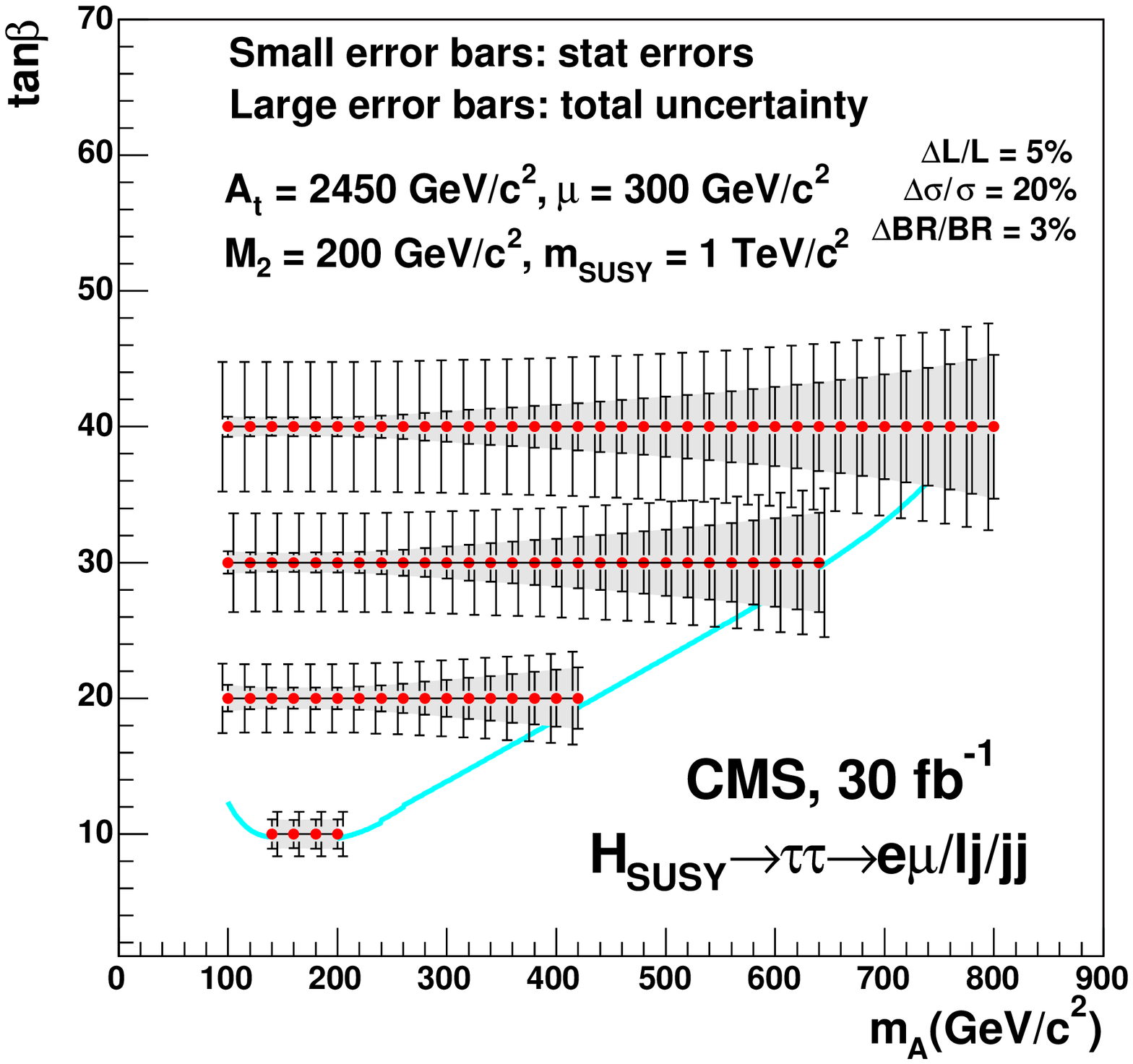}
  \caption{The uncertainty on the tan$\beta$ measurement shown as error bars.
           The small error bars and gray area show the statistical errors only.
           The large error bars show the total uncertainty evaluated as
           described in the text. 
            The solid curve corresponds to the 5$\sigma$-discovery contour of 
            Fig.~\ref{fig:discovery_HA}. Those results correspond to 
            30 fb$^{-1}$ of integrated luminosity.}
  \protect\label{fig:errorbars30fb}
  \centering
  \includegraphics[width=110mm]{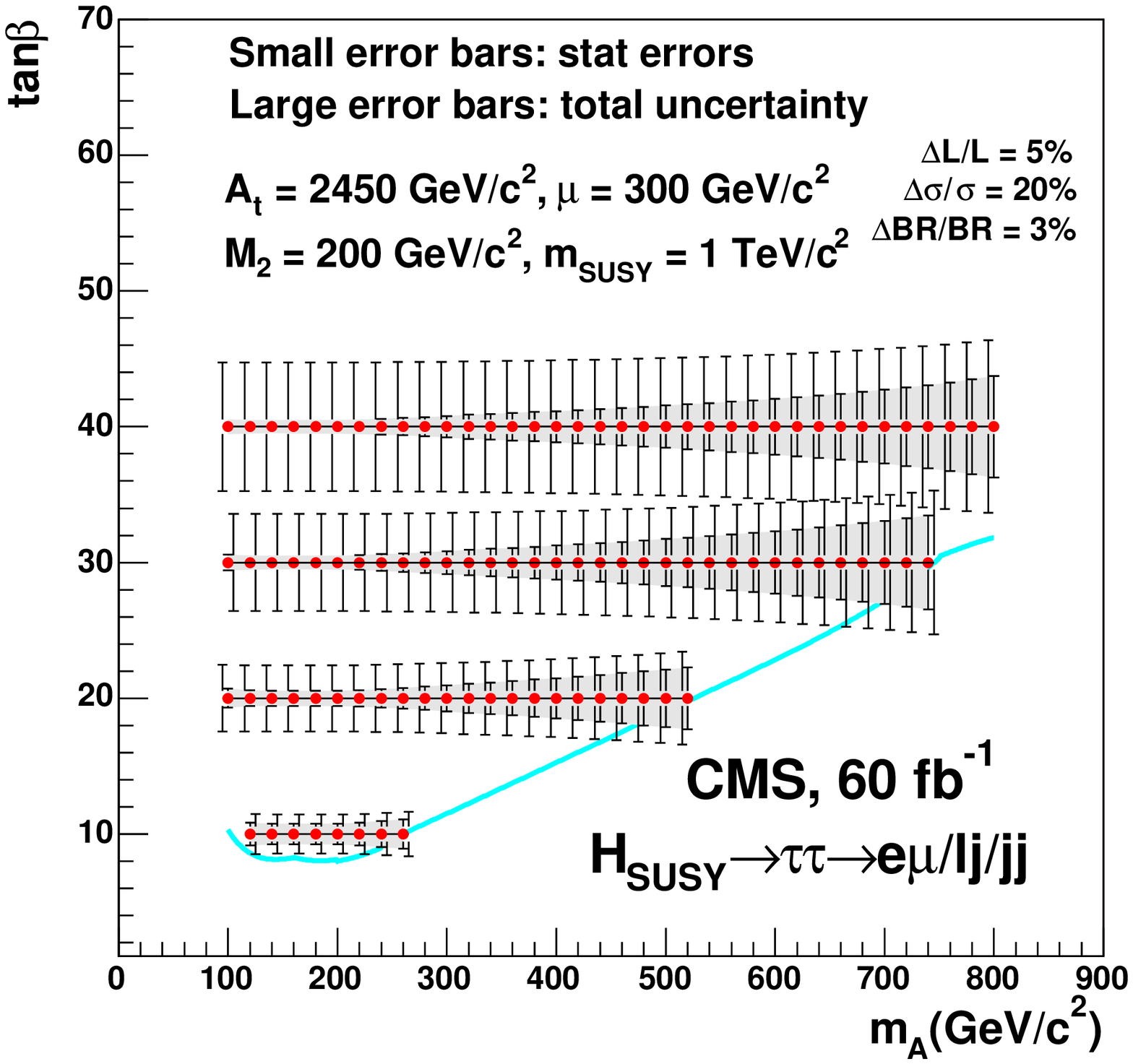}
  \caption{The same as in Fig.~\ref{fig:errorbars30fb} but with 
   60 fb$^{-1}$ taken at low luminosity.}
  \protect\label{fig:errorbars60fb}
\end{figure}

Figures \ref{fig:errorbars30fb} and \ref{fig:errorbars60fb} show the
uncertainty on the tan$\beta$ measurement with error bars for the combined
e$\mu$ + $\ell$j + jj channel for 30 and 60 fb$^{-1}$ at low luminosity.
The statistical uncertainties are depicted by the smaller error bars and
gray area, the uncertainties including the systematic errors are
presented with longer error bars. The errors are shown in the region
with signal significance larger than 5$\sigma$.  The combined 5$\sigma$
reach is plotted with the contribution from the e$\mu$ final state
included up to m$_{\rm A}$ = 180 GeV/$c^2$ in order to extend the
reach to lower tan$\beta$ values.  For the same reason at very high 
values of m$_{\rm A}$ only the two-jet final
state contributes to the reach. The errors are calculated within the
shown 5$\sigma$ reach using all available information, including
leptonic final states for m$_{\rm A}<$ 300 GeV/$c^2$, and
$\ell$j final state for m$_{\rm A}<$ 800 GeV/$c^2$.  
The statistical uncertainty is largest close to the
5$\sigma$ limit, where the combination of the different final states improves 
the accuracy.

\subsection{H/A $\ra\mu\mu$}

In the region m$_{\rm A} \lsim$~300~GeV/$c^2$, the value of tan$\beta$
could also be measured in the H/A$\rightarrow \mu\mu$ channel
using event rates. In this channel, the Higgs mass resolution is about
2\% \cite{bellucci}. Therefore the total width of the Higgs boson could
be measured with good precision from the Higgs boson mass peak.  The
variation of the natural width as a function of tan$\beta$, from less
than 1 GeV/$c^2$ to about 20 GeV/$c^2$ in the expected tan$\beta$ range, 
could be
used to determine the value of tan$\beta$. This method, based on the
direct width measurement, would therefore be complementary to the method
explained in this note.

\section{Conclusion}

The precision of the cross section times branching ratio, 
$\sigma ^{\rm prod}$,
measurement and the derived tan$\beta$ value was estimated in the
H/A~$\ra\tau\tau$ decay channel with two-lepton, lepton+jet and two-jet
final states for 30 fb$^{-1}$. The statistical precision on 
$\sigma ^{\rm prod}$ is expected to be 3$-$25 \% and the associated systematic 
error $\leq$ 12 \% both depending on the signal significance.

In the region of large tan$\beta$, the tan$^2
\beta$ dependence of the associated production processes $\rm gg \ra \rm b
\overline{\rm b} \rm H/A$ has been exploited to obtain
a statistical uncertainty being a factor of two smaller than that of the
event rates. Due to the presence of potentially large radiative
corrections to the bottom Yukawa coupling, the results
obtained in this analysis correspond to an effective parameter
tan$\beta_{eff}$ which absorbs the leading universal part of these
corrections. A theoretical error of 20\% (cross section) and 3\% 
(branching ratio) and a luminosity uncertainty of
5\% have been assumed.  If two b jets with $\rm E_{\rm T}>$~20~GeV are
tagged, the theoretical uncertainty on the cross section reduces to 
10-15\%, but the event
rates have been found to decrease significantly leading to a worse
accuracy of the tan$\beta$ measurement.
With one tagged b jet in the event the total uncertainty on tan$\beta$
estimated with the H/A~$\ra\tau\tau$ decay channels ranges from 12\% to 19\% 
within the 5$\sigma$-discovery reach depending on tan$\beta$ and 
$\rm M_{\rm A}$ after collecting 30~fb$^{-1}$.
The combination of the $\rm e\mu$, $\ell$+jet and two-jet or $\ell\ell$,
$\ell$+jet and two-jet final states gives 4\% better
statistical accuracy than the best individual final state.
Close to the 5$\sigma$-discovery limit the statistical uncertainty
ranges in the same order as the theoretical one, but for tan$\beta$
regions where the signal significance exceeds $5\sigma$ significantly
the theoretical error dominates.

\section{Acknowledgments}

The authors would like to thank Michael Kr\"amer, Fabio Maltoni, Tilman
Plehn, Torbj\"orn Sj\"ostrand, Georg Weiglein and Scott Willenbrock for 
helpful comments and discussions.

The authors would like to thank the members of the CMS Editorial Board 
Reyes Alemany Fernandez and Luc Pape for many useful comments which 
improved the paper.

%==============================================================================

\end{document}